\documentclass[fleqn,usenatbib]{mnras}

\usepackage{newtxtext,newtxmath}

\usepackage[T1]{fontenc}

\DeclareRobustCommand{\VAN}[3]{#2}
\let\VANthebibliography\thebibliography
\def\thebibliography{\DeclareRobustCommand{\VAN}[3]{##3}\VANthebibliography}

\usepackage{graphicx}	
\usepackage{amsmath}	
\usepackage{multirow}
\usepackage{url}
\usepackage{hyperref}
\usepackage{ae,aecompl}
\usepackage{threeparttable}
\usepackage{booktabs}

\usepackage{wasysym} 
\usepackage{scalerel} 
\usepackage{verbatim} 
\usepackage{xcolor}
\usepackage{listings}

\definecolor{grey}{rgb}{0.4,0.5,0.6}
\definecolor{brown}{rgb}{0.65,0.16,0.16}
\definecolor{darkgreen}{rgb}{0.0,0.45,0.0}
\definecolor{darkorange}{rgb}{0.9,0.2,0.0}

\defcitealias{DiazGimenez+20}{Paper~I}
\defcitealias{DiazGimenez+21}{Paper~II}
\defcitealias{Taverna+22}{Paper~III}
\defcitealias{Taverna+22r}{III}
\defcitealias{zandivarez+23}{Paper~IV}

\title[CGs assembly channels and their environments]{Compact groups from semi-analytical models of galaxy formation -- V:\\
   their assembly channels as a function of the environment}

\author[A. Taverna et al.]
{A. Taverna$^{1,2,3}$\thanks{ataverna@unc.edu.ar}, 
E. D\'iaz-Gim\'enez$^{1,2}$, A. Zandivarez$^{1,2}$, H.J. Mart\'inez$^{1,2}$, A.N. Ruiz$^{1,2}$
\\
$^{1}$ Universidad Nacional de C\'ordoba (UNC). Observatorio Astron\'omico de C\'ordoba (OAC). C\'ordoba, Argentina\\
$^{2}$ CONICET. Instituto de Astronom\'ia Te\'orica y Experimental (IATE). C\'ordoba, Argentina\\
$^{3}$ Instituto de Astronomía, UNAM, Apdo. Postal 106, Ensenada 22800, B.C., México\\
}

\date{Accepted XXX. Received YYY; in original form ZZZ}

\pubyear{2023}

\begin{document}
\label{firstpage}
\pagerange{\pageref{firstpage}--\pageref{lastpage}}
\maketitle

\begin{abstract}
We delved into the assembly pathways and environments of compact groups (CGs) of galaxies using mock catalogues generated from semi-analytical models (SAMs) on the Millennium simulation. 
We investigate the ability of SAMs to replicate the observed CG environments and whether CGs with different assembly histories tend to inhabit specific cosmic environments. We also analyse whether the environment or the assembly history is more important in tailoring CG properties. 
We find that about half of the CGs in SAMs are non-embedded systems, 40\% are inhabiting loose groups or nodes of filaments, while the rest distribute evenly in filaments and voids, in agreement with observations. 
We observe that early-assembled CGs preferentially inhabit large galaxy systems ($\sim 60\%$), while around 30\% remain non-embedded. Conversely, lately-formed CGs exhibit the opposite trend. 
We also obtain that lately-formed CGs have lower velocity dispersions and larger crossing times than early-formed CGs, but mainly because they are preferentially non-embedded. Those lately-formed CGs that inhabit large systems do not show the same features. Therefore, the environment plays a strong role in these properties for lately-formed CGs.
Early-formed CGs are more evolved, displaying larger velocity dispersions, shorter crossing times, and more dominant first-ranked galaxies, regardless of the environment. 
Finally, the difference in brightness between the two brightest members of CGs is dependent only on the assembly history and not on the environment.
CGs residing in diverse environments have undergone varied assembly processes, making them suitable for studying their evolution and the interplay of nature and nurture on their traits.
\end{abstract}




\section{Introduction}
\label{sec:intro}
Compact groups of galaxies (hereafter CGs) have interested the astronomical community mostly since their first systematic observational searches \citep{Shakhbazyan73, Petrosian74, Rose77}. Their study probably intensified after the work of \cite{Hickson82}, where a new catalogue of CGs was released identified visually on the plates of the Palomar Observatory Sky Survey, and a set of criteria was proposed to identify them in observational catalogues. This encouraged the construction of new observational catalogues via automatic search algorithms most of them based on the Hickson criteria (e.g. \citealt{Prandoni+94,McConnachie+09,DiazGimenez+12,sohn+15,DiazGimenez+18}).  

By studying CGs, it is possible to gain valuable insights into the mechanisms driving galaxy evolution. Since they consist of a small number of galaxies in close proximity, they are assumed to be suitable places in the Universe to study the intricate dynamics and interactions between galaxies. Tidal interactions, mergers and close encounters are among the processes that are expected to occur frequently in CGs and play an important role in shaping galaxy properties. However, to grasp the real nature of such a very dense environment described by CGs, a full understanding of their place in the Universe is crucial. 

Over the last thirty years, many studies have explored the environment surrounding CGs. The results varied, with a wide range of percentages showing CGs embedded in larger systems ($\sim 10\%$ - \citealt{sohn+15}; $20-30 \%$ -  \citealt{palumbo+95,decarvalho+05,DiazGimenez+15}; $\sim 50\%$ - \citealt{andernach+05,mendel+11,zheng+21}; $\sim 70\%$ - \citealt{rood+94,barton+98}). Using several mock catalogues, \cite{Taverna+22} observed that roughly $90\%$ of a sample of Hickson-like CGs is most likely embedded in a larger galaxy system in 3D real space.

In recent work, \cite{taverna+23} used Hickson-like CGs identified in the Sloan Digital Sky Survey Data Release 16 (SDSS DR16, \citealt{sdss_dr16}) to analyse when they can be considered embedded or not in different cosmological structures. They observed that 45 per cent of CGs can be considered non-embedded or isolated systems. At the same time, most of the embedded CGs were found inhabiting Loose Groups and Nodes of filaments, and very few residing in Filaments or the surroundings of cosmic Voids. 
\cite{taverna+23} also observed that CGs in Nodes show the largest velocity dispersions, the brightest first-ranked galaxy, and the shortest crossing times. In contrast, the opposite characteristics are observed in Non-Embedded CGs.

Recently, using several mock catalogues built from numerical simulations and semi-analytical models of galaxy formation (SAMs), we performed a series of works that have been carried out trying to study CGs in great detail. 
Analysing the frequency and nature of Hickson-like CGs, in \cite{DiazGimenez+20} (hereafter \citetalias{DiazGimenez+20}) we discovered a dependence on the cosmological parameters and SAM recipes. In \cite*{DiazGimenez+21} (hereafter \citetalias{DiazGimenez+21}) from the study of the galactic orbits of Hickson-like CG galaxy members, we introduced a new classification of CGs based on their assembly history. Discarding those CGs that are not compact in 3D real space, our definition of four different assembly channels has shown that most CGs ($\sim 75$ per cent) assembled recently (with the latest galaxy to arrive having done just one or two close approaches). About 10 to 20 per cent of CGs are better described by a gradual contraction during the last 7 Gyrs while the remaining few ($\sim 3-8$ per cent) of them can be considered systems of early assembly. We also explored the performance of the Hickson-like automatic algorithm used in flux-limited catalogues in \cite{Taverna+22} (hereafter \citetalias{Taverna+22}), finding that this algorithm performs poorly to recover isolated dense groups identified in 3D real space (mainly because they are not truly isolated in 3D). Finally, in \cite{zandivarez+23} (hereafter \citetalias{zandivarez+23}) we analysed the time evolution of the properties of galaxies in CG as a function of their assembly channel. We have observed that galaxies residing in early assembled CGs are more efficient at consuming gas, growing their black holes, and suppressing their star formation, compared to what is observed for galaxies inhabiting recently formed CGs.

In this work, we extend the work of \cite{taverna+23} related to the location of CGs in several environments, this time within the framework of the series of works on CGs identified in different SAMs of galaxy formation described previously. Our aim is two-fold: first, we analyse whether the distribution of CGs in different galactic structures, now identified in mock lightcones constructed from several SAMs, resembles the results obtained by \cite{taverna+23} in observations; and second, we use the classification of CGs based on their assembly history (\citetalias{DiazGimenez+21}) to understand if there is a particular environment where CGs of a given evolutionary history can be found.

This study used four mock CG samples built from the Millennium Simulations \citep{Springel+05}, with SAMs used in previous works (I, III and IV). 
The different assembly channels of CGs have been defined using the procedure proposed in \citetalias{DiazGimenez+21}. 

The layout of this work is as follows. In Sect.~\ref{sec:samples}, we present the mock lightcones built from different SAMs, and the mock CG and galaxy systems used in this work. 
In Sect.~\ref{sec:structures} we compare our results in mock catalogues with those obtained by \cite{taverna+23} in an observational sample. In Sect.~\ref{sec:channels}, we describe the classification of CGs according to their assembly channel, while in Sect.~\ref{sec:results} we analysed the preferred environment inhabited by CGs as a function of the assembly channel. We conclude and discuss our results in Sect.~\ref{sec:conclusions}.

\begin{table}
\caption{Number of galaxies and galaxy systems in each mock lightcone.
}
\label{tab:mock}
\begin{tabular}{rrrrrr}
\hline
\multicolumn{1}{l}{}           & \multicolumn{1}{c}{G11} & \multicolumn{1}{c}{G13} & \multicolumn{1}{c}{H20} & \multicolumn{1}{c}{A21} \\
\hline
mock galaxies & 2977881   & 2809956   & 3172243  & 2724272 \\
 \hline
CGs & 7850      & 5233      & 5881     & 2927  \\
FoF  & 66135     & 61408     & 71974    & 39790  \\
Nodes  & 1988 & 1868 & 1826 & 372 \\   
Filaments  & 1945 & 1852 & 1772 & 287\\
Voids $V_{0.1}$  & 1863 &  1970 &  2001 & 1759 \\
LGs &  61655    &  57962   &   68365  & 38526 \\
\hline
\hline
CG$_{0.1}$       & 7410 & 4969 & 5509 & 2757 \\
CG$^{3+4}$       & 7129 & 4901 & 5487 & 2794 \\
CG$^{3+4}_{0.1}$ & 6713 & 4648 & 5126 & 2630 \\
\hline
\end{tabular}
\parbox{\hsize}{Notes:mock galaxies with $r\le17.77$ \& $\mathcal{M}^\star \ge 7\times 10^8 \, \mathcal{M}_\odot \, h^{-1}$ \& $z_{\rm max}=0.2$. 
CGs: compact groups with three or more observable$^{(*)}$ members. 
FoF: normal groups with four or more observable members and virial masses larger than $10^{12} \mathcal{M}_\odot \, h^{-1}$. 
Nodes: normal groups with ten or more observable members and virial masses larger than the median virial mass that connect filaments. 
Filaments: overdense cylinders with then or more observable members. 
Voids: underdense spheres with 80\% or more of their volume within $V_{0.1}$.
LGs: FoF groups that are not Nodes nor CGs. 
CG$^{3+4}$: compact groups that are triplets or quartets.
CG$_{0.1}$: compact groups within the volume-limited $V_{0.1}$.
CG$^{3+4}_{0.1}$: triplets and quartets within $V_{0.1}$.
(*) The word observable is included to distinguish between the real number of members and the number of members that could be potentially observed considering the blending of galaxies.
}
\end{table}
\section{Samples}
\label{sec:samples}

\subsection{Mock galaxy Lightcones}
As in the previous papers of this series, we build four lightcone catalogues by using the several outputs of the Millennium I Simulation \citep{Springel+05, Guo+13} with different cosmologies and combined with the following four SAMs that are publicly available\footnote{\url{http://gavo.mpa-garching.mpg.de/Millennium/}}: G11 \citep{Guo+11}, G13 \citep{Guo+13}, H20 \citep{Henriques+20} and A21 \citep{Ayromlou21}. 
The lightcones are built following the procedure detailed in \cite{jpas}. The evolution of galaxy properties and structures is included by using galaxies from different snapshots in the simulations. Observer-frame apparent magnitudes are computed from the rest-frame absolute magnitudes provided by the SAMs and a K-decorrection procedure is applied \citep{DiazGimenez+18}. 
The mock lightcones span the whole sky and comprise galaxies with stellar masses greater than $7\times 10^8 \, \mathcal{M}_\odot \, h^{-1}$ 
($\sim 10^9 \mathcal{M}_\odot$), redshifts less than 0.2, and r-band observer frame apparent magnitudes less than 17.77 to mimic the flux limit of the SDSS. In Table~\ref{tab:mock} we quote the number of galaxies in each catalogue. 

\subsection{Galaxy Systems}
The galaxy systems used in this work are identified with the same criteria and algorithms used in \cite{taverna+23} on the SDSS DR16, this time applied to the mock lightcones. In the following sections, we briefly describe each sample but for more details, we recommend the reading of \cite{taverna+23}.

\subsubsection{CG samples}
\label{sec:CG_iden}
We identify CGs that accomplish the following observational criteria: 
\begin{description}
    \item[Population:] There are at least three galaxy members. All the galaxy members are within a three-magnitude range from the brightest galaxy of the system. 
    \item[Flux limited:] The brightest galaxy of the group is three magnitudes brighter than the magnitude limit of the lightcones.      
    \item[Velocity concordance:] All the galaxy members are within $1\,000 \rm \, km \, s^{-1}$ from the median radial velocity of the group.
    \item[Relative isolation:] The group is relatively isolated, that is: there are no other bright galaxies (3-mag range) within a disk of three times the minimum circle that encloses the galaxy members.  
    \item[Compactness:] The surface brightness of the group in the r-band is brighter than $26.33 \, \rm mag \, arcsec^{-2}$.
\end{description}

We include the possible blending of galaxies that exists in observations by assigning a half-light radius to each dimensionless mock galaxy member. Therefore, we consider two mock galaxies as blended if their radii overlap in projection. The half-light radii are assigned as a function of the galaxy stellar mass in the SAMs by following the prescriptions of \cite{Lange+15}. The blending only affects the memberships, although it might be important in triplets since one pair of galaxies blended makes the membership drop to 2 and the group is discarded. In what follows, we will refer to the number of members after blending as \textit{observable} number of members. The fractions of identified CGs that survive after the blending procedure (3 or more observable members) are 0.72, 0.78, 0.71 and 0.85 for G11, G13, H20 and A21, respectively.
The number of CGs (after blending) identified in each mock lightcone is quoted in  Table~\ref{tab:mock}.
 
\subsubsection{Galaxy Group samples}
\label{sec:FoF_iden}
We identify galaxy groups by means of a Friends-of-Friends algorithm in redshift space \citep{Huchra82}. The linking-length parameters used in this work are the same used by \cite{zandivarez+22} in the SDSS DR16 catalogue, but taking into account the different cosmologies used in the simulations. The values of the transversal and line-of-sight linking length $(\rm d0 [kpc/h],v0 [km/s])$ are $(226,261)$ in G11, $(231,269)$ in G13, and $(238,285)$ en H20 and A21. 

We include the blending of galaxies as we described in the previous section for CGs, and then we select normal groups with four or more observable members and virial masses larger than $10^{12} {\cal M}_\odot \, h^{-1}$. The number of groups in each lightcone is quoted in the third row of Table~\ref{tab:mock}, and we will refer to this sample as the FoF groups.

\subsubsection{Filament and Node samples}
\label{sec:Fila_iden}
Filaments are overdense cylindrical volumes in redshift space that connect rich massive groups denoted as nodes. 
The algorithm starts with a sample of potential nodes (FoF groups with virial masses larger than the median mass of the sample of FoF groups and having ten or more observable galaxy members) and looks for pairs of groups separated by less than a threshold distance (in redshift space and projected sky), and computes the overdensity of galaxies within a cylinder of fixed radius that connects both groups. 
When the overdensity surpasses a given threshold and the number of observable galaxies within the cylinder is larger than ten, then the filament and its nodes are saved.
The numbers of filaments and nodes identified in each lightcone are quoted in Table~\ref{tab:mock}.

\subsubsection{Loose Groups}
\label{sec:LG_iden}
From the sample of FoF groups, we discard those FoF groups that have already been identified also as CGs\footnote{The procedure to match FoF groups with CGs is not a member-to-member comparison, but instead it takes into account the restrictions in flux limit and isolation required by the CG searching algorithm. The matching FoF groups have to include 75\% of the CG members, while the number of FoF bright members outside the CG isolation disk cannot surpass half the number of CG members.}, which is less than $4\%$ 
of the FoF samples, and also those FoF groups that have been classified as Nodes of filaments. 
The remaining sample of FoF groups is referred to as Loose Groups (LGs) in Table~\ref{tab:mock}. 

\subsubsection{Void samples}
\label{sec:void_iden}
To identify Voids, a volume-limited sample is defined by selecting galaxies with r-band absolute magnitude brighter than $-19.769$ and redshift less than $0.1$. This volume-limited sample will be referred to as $V_{0.1}$. 
As in \cite{taverna+23}, cosmic voids are identified as under-dense spheres in the volume-limited sample. 
Afterwards, we include as voids members all the fainter galaxies in the flux-limited sample that lie within the sphere.  
The number of voids identified in each SAM is quoted in Table~\ref{tab:mock}. 

\begin{figure}
    \centering
    \includegraphics[width=\columnwidth, trim=10 15 18 0, clip]{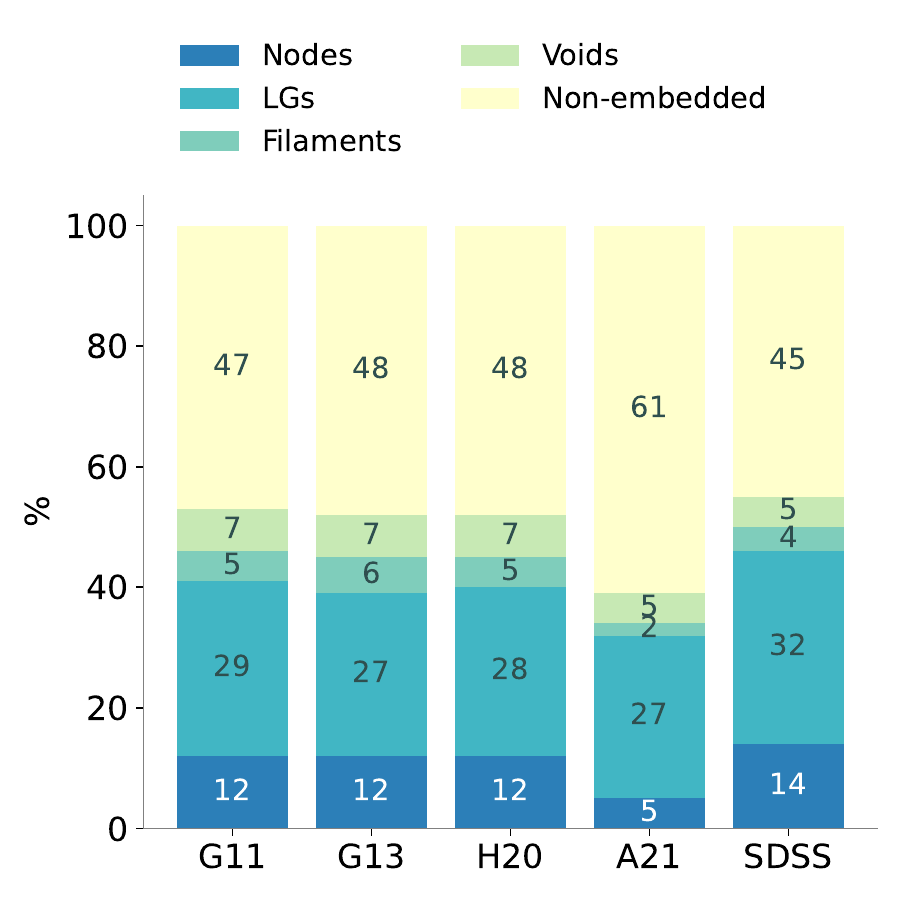}
    \caption{Percentage of CGs in different cosmological structures for each SAM. In the last bar, we quote the percentages obtained by \protect\cite{taverna+23} for the observational sample of CGs identified in the SDSS DR16 survey \protect\citep{zandivarez+22}.}
    \label{fig:cg_struct}
\end{figure}
\section{Location of CGs in the cosmic web}
\label{sec:structures}
By definition, Hickson-like criteria identify CGs that are locally and relatively isolated, since the criterion is based on the projected size of the CG and the relative brightness of the neighbouring galaxies. 
Therefore, the isolation criterion does not prevent a CG from being inside a larger structure, such as filaments, voids, nodes or loose groups, as it was demonstrated in \cite{taverna+23}. 

To look for CGs embedded in larger structures, in this section, we restrict the sample of CGs to systems that lie within the volume-limited sample defined for the voids identification procedure. That is, the r-band absolute magnitude of the brightest galaxy is less than~$-19.769$ and the median redshift of the CG is less than~$0.1$. The number of CGs within $V_{0.1}$ in each mock lightcone is quoted in Table~\ref{tab:mock} (referred to as CG$_{0.1})$.

To determine if the CGs are associated with a larger structure, we perform the same procedure that \cite{taverna+23} (see details in their Section 3).  
We can summarise the process in the following two steps: 
\begin{itemize}
    \item a member-to-member comparison: we consider a CG associated with Nodes, Filaments or Voids if they share at least one CG member, or two members with an LG. 
    \item location of the CG centre: we also consider a CG associated with a Filament if the CG centre is inside the cylindrical region defined by the Filament, and it is considered as associated with a Void if the CG centre is within the comoving sphere of $1.1$ times the Void radius. 
\end{itemize}

Finally, according to the CG associations performed in the previous two steps, we classify each CG as CGs in Nodes, CGs in Filaments, CGs in Loose Groups, and CGs in Voids. If a CG is not associated with any of these structures, we consider the CG as non-embedded. Although the sample of non-embedded CGs is not associated with any previously identified structures, they are not necessarily locally isolated systems. They could be linked to galaxy pairs or triplets, or even to groups with lower overdensity contrast than our FoF groups, or they may also be associated with filaments with fewer than 10 members.

In Fig.~\ref{fig:cg_struct}, we show 
the percentage of CGs inhabiting different environments for each SAM. In the last column, we show the results obtained by \cite{taverna+23} for the observational sample of CGs identified in the SDSS-DR16. 
We find that most CGs are non-embedded systems (around 50\%), with a high percentage living in loose groups (around 28\%). Roughly 12\% of CGs inhabit nodes of filaments, while around 5\% live in filaments. The remaining CGs are within Voids (around 5\%). 
A21 differs the most, displaying almost half of the CGs in Nodes than the other SAMs, and approximately 10\% more non-embedded CGs than the other SAMs. 
Despite these differences, the percentages of CGs from SAMs inside each structure broadly agree with the percentages found for the observational CG sample by \cite{taverna+23}. 
This result implies that the mock catalogues build from these SAMs are able to replicate the same type of environments that surround CGs in observations. Perhaps, the lower percentage of CGs in Nodes for A21 is a consequence of stronger recipes in this SAM to deal with environmental processes. In \citetalias{zandivarez+23} we observed that A21 has a lower fraction of orphan galaxies (satellites without a dark matter subhalo) than those observed in the other SAMs. This could be a result of A21 implementing more efficient tidal and ram-pressure stripping in dense environments since they were implemented using the local background environment measurements. This could imply a lower rate of survival of CGs in very dense environments such as Nodes.

\begin{figure}
    \begin{center}
    \includegraphics[width=0.95\hsize]{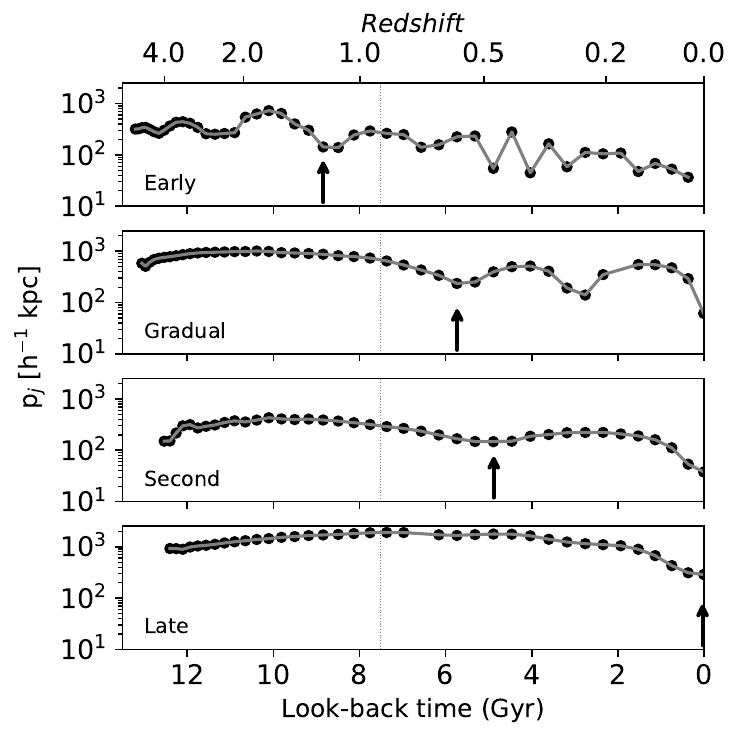}
    \caption{3D physical distance of the key galaxy member to the CG centre of mass as a function of look-back time. These are four examples of CGs with different assembly channels for the A21 SAM.
    (from \emph{top} to \emph{bottom}: {\tt Early}, {\tt Gradual}, {\tt Second} and {\tt Late}). Vertical lines indicate the look-back time (7.5 Gyr) to separate Early assembled CG from the other three channels.Arrows indicate the 1st deep pericenter of each key galaxy.
    \label{fig:orbits}}
    \end{center}
\end{figure}
\begin{figure*}
    \begin{center}
    \includegraphics[width=0.99\hsize]{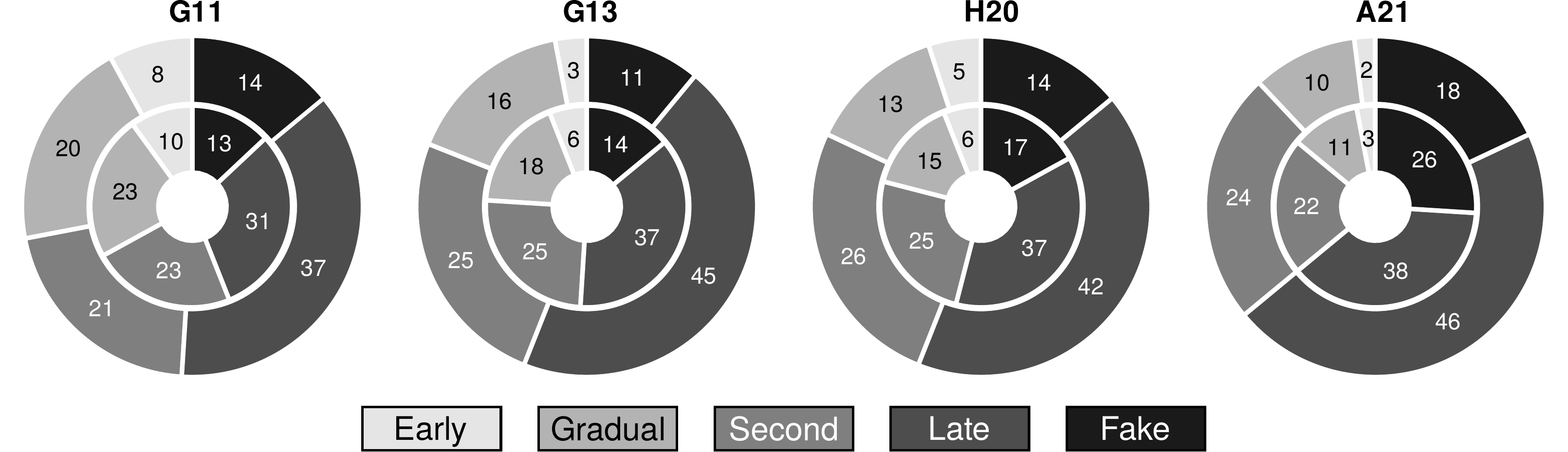}
    \caption{Ring diagrams displaying the percentage of CGs in each assembly channel for the different SAMs used in this work (ordered from earliest to latest assembled progressing with lighter to darker shades of grey). We also include the percentage of CGs that can be considered false detections labelled as Fake (in black). The outer ring quotes the percentages of CG samples with three and four members (CG$_{3+4}$), while the inner ring shows CGs formed by only four galaxies. The assembly channels in the ring diagrams are oriented counterclockwise, ordered from earliest to latest (plus Fake is placed at the last position).
    \label{fig:donut}}
    \end{center}
\end{figure*}
\section{Assembly channels of CGs}
\label{sec:channels}

The classification of CGs into different assembly channels follows a specific procedure outlined by \cite{DiazGimenez+21} (referred to as \citetalias{DiazGimenez+21}). The process involves tracking the merger tree histories
of the main progenitor of the individual CG members at redshift $z=0$ to determine their evolutionary paths.

For each galaxy member $j$, we determine its physical 3D distance to the stellar mass centre of the group in each output $i$ of the simulation\footnote{$i$ runs from 0 at $z=0$ to the total number of snapshots in each SAM}, $p_j(t_i)$, where $t_i$ is the lookback time of the snapshot $i$. From this profile, we measure the time of the galaxy's first pericenter passage ($t_{\rm 1p}$), the number of (deep\footnote{A deep pericenter at $t_i$ is defined when the height of the 3D physical distance profile is less than 0.8 of the height at $t_{i-2}$ and $t_{i+2}$.}) pericentres passages ($n_{\rm p}$), and the ratio between $p_j(t_i)$ in $t_0$ (snapshot at $z=0$) and the immediately previous output $t_1$ ($p_j(t_0)/p_j(t_1)$).
As detailed in Sec.~4.2 of \citetalias{DiazGimenez+21}, four different types of CG assembly channels are defined based on the behaviour of the galaxy member with the latest arrival to the group, referred to as the "key" galaxy\footnote{The "key" galaxy is defined as the one with the lowest $n_{\rm p}$ value. When several galaxies have the same lowest $n_{\rm p}$, the one with the most recent $t_{\rm 1p}$ is selected, provided that their $t_{\rm 1p}$ values are separated by more than 1 Gyr. If this condition is not met for any galaxy, the "key" galaxy is chosen based on the highest value of $p_j(t_{\rm 1p})$ at its first pericenter.}:

\begin{itemize}
    \item If the key galaxy has joined the CG at late times, i.e. $t_{\rm 1p} \le 7.5$ Gyr, then there are three possibilities:
    \begin{itemize}
        \itemindent=0pt
        \item  {\tt Late Assembly}: the fourth galaxy has just arrived in the CG for the first time (hereafter {\tt Late}). This happens when $n_{\rm p}=0$ OR [$n_{\rm p}=1$ AND $p_{\rm key}(t_0)/p_{\rm key}(t_1)>1$].
        \item {\tt Late Second Pericentre}: the fourth galaxy has arrived in the CG on its second passage (hereafter {\tt Second}), i.e. when [$n_{\rm p}=1$ AND $p_{\rm key}(t_0)/p_{\rm key}(t_1)\leq 1$] OR [$n_{\rm p}=2$ AND $p_{\rm key}(t_0)/p_{\rm key}(t_1) > 1$].
        \item {\tt Gradual Contraction}: the four galaxies have already completed two or more orbits together, becoming gradually closer with each orbit. This behaviour (hereafter {\tt Gradual}) is obtained when [$n_{\rm p}=2$ AND $p_{\rm key}(t_0)/p_{\rm key}(t_1) \leq 1$] OR $n_{\rm p}>2$.
    \end{itemize}
    \item If the key galaxy has an early arrival at the CG, i.e. $ t_{\rm 1p} \ge 7.5$ Gyr:
    \begin{itemize}
        \itemindent=0pt
        \item {\tt Early Assembly}: the four galaxies have been together since an early epoch (hereafter {\tt Early}).
    \end{itemize}
\end{itemize}
 
In Fig.~\ref{fig:orbits}, we display four examples of the key galaxy orbits around the stellar mass centre to illustrate the idea behind the four assembly channel definitions. 

In \citetalias{DiazGimenez+21}, the classification was applied only for CGs with exactly four bright galaxy members, the lowest membership of the sample used in that work which represented $\sim 90\%$ of the total sample. 
CGs with higher membership were discarded to simplify the study of the galaxy orbits for each CG. 
In this work, since we are interested in performing a prediction about assembly channels for the observational results obtained by \cite{taverna+23}, we include in our analysis not only those CGs with four members but also CGs with three galaxy members. In this manner, we are also including in this work more than $90\%$ of the mock CG samples. 
The number of CGs with only three and four members (CG$^{3+4}$) is quoted in Table~\ref{tab:mock}.
The definition of assembly channels remains the same of \citetalias{DiazGimenez+21}, but for triplets, the key galaxy is the third to arrive in the systems. 

It is worth noticing that, following \citetalias{DiazGimenez+21}, we use the real-space information to exclude from our classification those CGs with maximum 3D inter-galaxy separations larger than $1 \ h^{-1}$ Mpc at $z=0$, and those without close pairs in 3D space\footnote{Close galaxy pairs are those whose galaxies are separated by less than $200 \ h^{-1}$ kpc.}. These systems, which are clearly chance alignments along the line of sight, are labelled as "Fake" CGs in the subsequent parts of this work and are not taken into account in the assembly channel classification, but instead are presented as a different subsample.

We summarize our results for the classification according to the assembly channel in Fig.~\ref{fig:donut}. These ring diagrams display the percentage of CGs in each assembly channel and also the percentage of Fake CGs. The outer rings are the percentages for the sample of CGs with three and four bright galaxy members, while the inner rings are the samples restricted to only quarters. 

From the samples of triplets and quartets (outer rings), Fake systems represent $14\%$ of the CGs samples. In addition, $\sim 43\%$ of CGs are Late assembled, roughly a quarter of CGs ($\sim 24\%$) are classified as Second, while $\sim 15\%$ are Gradual. The lowest percentage corresponds to Early assembled CGs with between $2$ to $8\%$ of CGs. 
Therefore, the general tendency is to have roughly two-thirds of the CGs represented by recently assembled systems (Late+Second).
Analysing differences from SAM to SAM, G11 shows the lowest fraction of recently assembled CGs (Late+Second $\sim 58\%$) and the highest fraction of gradual contraction and Early assembled systems (28\%), while the opposite happens in A21 (70\% and 12\% of recently and Gradual+Early assembled, respectively). 
Taking into account that the differences between G11 and G13 are mainly due to cosmology, and between H20 and A21 are primarily due to different SAM recipes, it is evident that both cosmology and SAM recipes play a significant role in the assembly channels of CGs. 

The results for the samples of only quartets are included for comparison purposes in the inner rings of Fig.~\ref{fig:donut}, given that in previous work only quartets were considered to perform the classification. 
In this case, the number of quartets represents a low percentage ($\sim 20\%$) of the CG$^{3+4}$ sample: 1526, 919, 1139, and 463, for G11, G13, H20 and A21, respectively.
Firstly, the results shown in these inner rings are in full agreement with the results presented in \citetalias{DiazGimenez+21} and \citetalias{zandivarez+23}. 
Secondly, from the comparison with the outer rings, we find that in the samples of triplets+quartets, there is an increase of $\sim 7\%$ of CGs with late assembly channel compared to the quartets-only samples, at the expense of Gradual ($-2\%$), Early ($-1\%$) and Fake CGs ($-3\%$, much notorious in A21 with $-8\%$). The percentage of CGs classified as Second remain the same. 
Despite these small differences, including triplets in the samples has not modified the general results presented in previous works: most CGs have been recently assembled, while very few are early assembled.

\begin{figure*}
    \begin{center}
    \includegraphics[width=0.95\hsize]{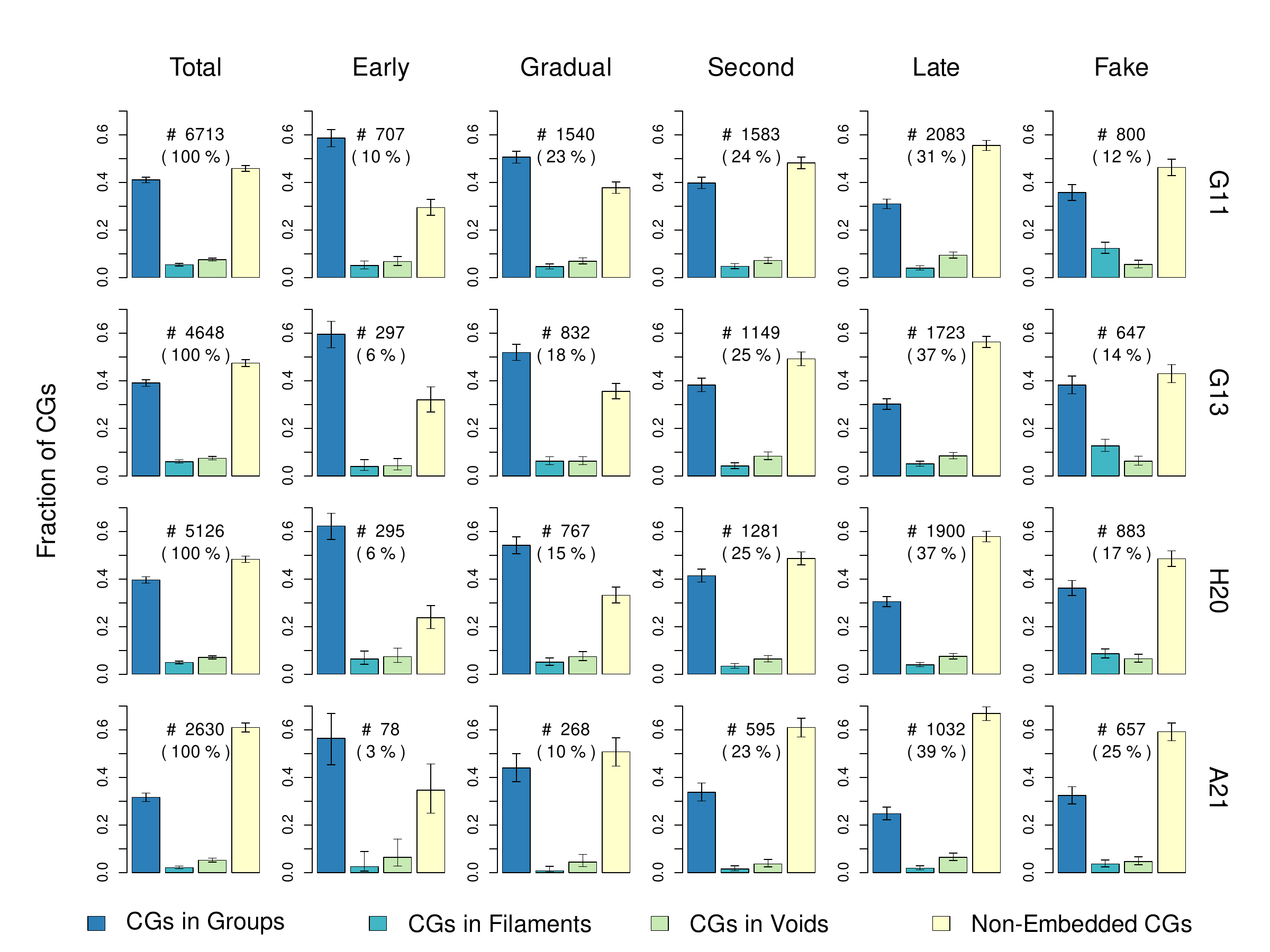}
    \caption{\label{fig:barplot} Fractions of CGs inhabiting different structures: groups, filaments, voids, and those considered Non-Embedded CGs. The first column corresponds to the total sample of CGs, while the other columns show the fractions for CGs with different assembly channels: early assembly, gradual contraction, late second pericentre, and late assembly. The last column corresponds to the remaining CGs not classified into formation channels since they are considered fake associations in real space. Each row shows the fractions for a given SAM.}
    \end{center}
\end{figure*}

\section{CG location in the universe as a function of their assembly channel}
\label{sec:results}
In this section we combine the information obtained in the previous two sections to answer the following question: do CGs with a particular history of assembling have a preferred place in the Universe? 

First, we select a subsample of CGs within the volume-limited sample $V_{0.1}$ (restriction necessary for the structure association) which comprises only triplets and quartets (restriction needed to classify into assembly channels). The number of CGs in this subsample is quoted in the last row of Table~\ref{tab:mock} (CG$^{3+4}_{0.1}$).
This subsample of CGs is then split according to the environment they inhabit (as outlined in Sec.~\ref{sec:structures}). For a more straightforward understanding of the various environments, we reduced the number of categories to four: 
CGs within Groups (includes CGs within Nodes and within Loose Groups), CGs within Filaments, CGs in Voids, and Non-Embedded CGs. Afterward, CGs within each environment are classified according to their assembly channel. 

The fractions of CGs of a given assembly channel in the different environments are displayed in Fig.~\ref{fig:barplot}. In this plot, we include the number of CGs in each assembly channel and the percentage they represent from the total sample shown in the first column. 
Regardless of the SAM, Early assembled CGs are typically located inside larger galaxy groups \textbf{($\sim 0.6$)}, roughly a third are non-embedded CGs, while the remaining sample \textbf{($\sim 0.1$)} are located either in filaments or cosmological voids. 
Quite the opposite is observed for CGs classified as Late since they are predominantly non-embedded systems ($\sim 0.6\%$), while approximately a third of them are embedded in larger galaxy groups. In the middle, CGs classified as Gradual or Second look like transitional stages between Early and Late CGs, with the Gradual CGs being more similar to Earlies while the Seconds are more similar to Late CGs. 
Regarding CGs classified as Fake systems (6th column), they behave similarly to the Total sample of CGs (1st column) with a small predominance outside galaxy systems, although Fake CGs in filaments are more frequent than in voids, contrary to what happens in the total sample. 

\begin{figure*}
    \begin{center}
    \includegraphics[width=0.95\hsize]{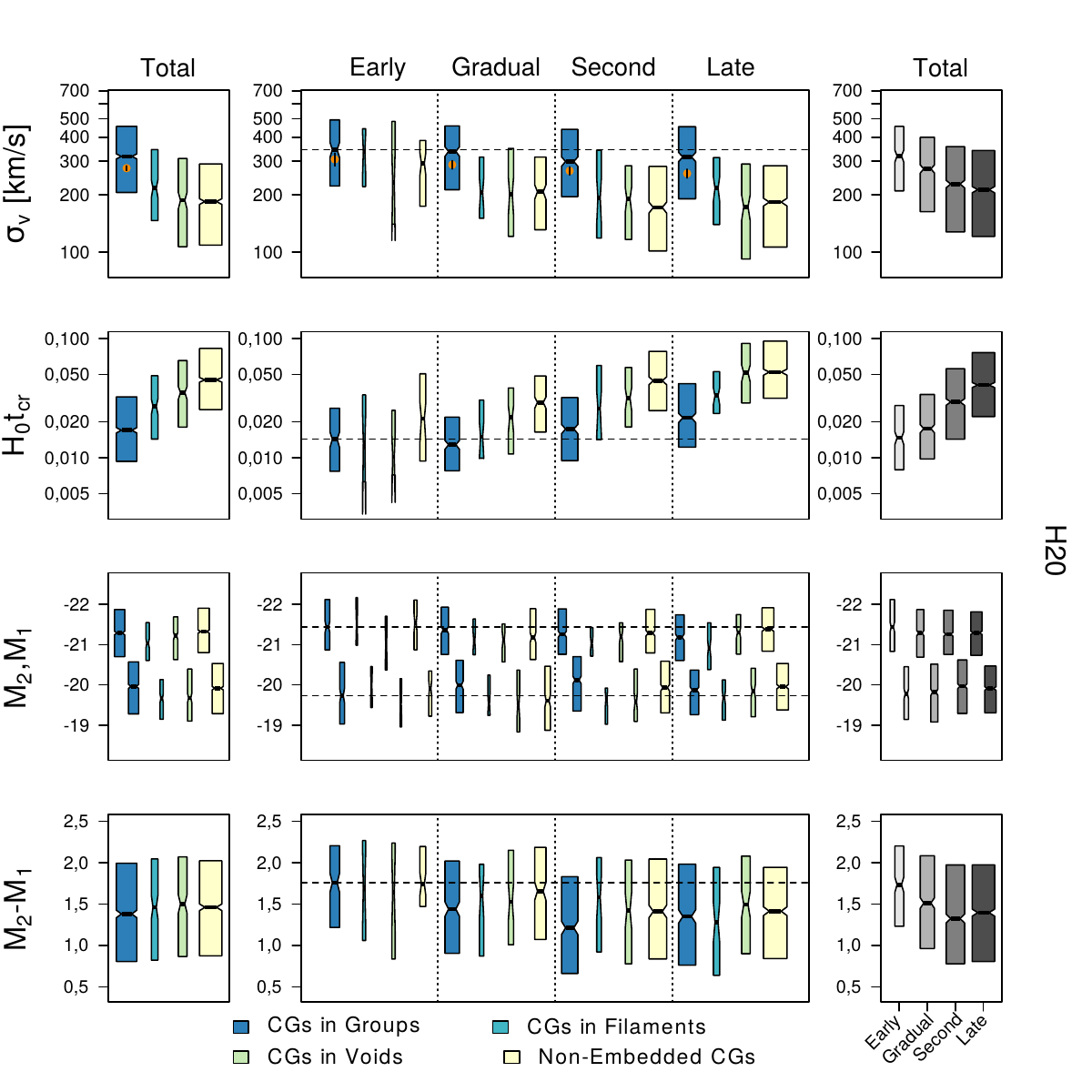}
    \caption{\label{fig:boxplot_h20} Boxplot diagrams of CG properties identified in H20 as a function of the environments and the assembly channel. Each box extends from the lower quartile (25th percentile) to the upper quartile (75th percentile). The notches indicate the approximate 95 percent confidence interval for the medians. The first column displays the boxplot diagrams of the Total sample as a function of the environment, while the last column is also for the Total sample but as a function of the assembly channel. The horizontal dashed line in the middle panels is only included as a reference for comparison between the samples and is set to the median value of the subsample of CGs in Groups with Early assembly channels. Solid orange dots in the first row indicate the median velocity dispersions of the sample of groups (loose groups and nodes) hosting CGs, while black segments on top of the dots represent the confidence intervals for these medians.}
    \end{center}
\end{figure*}

To complement our analysis, we also study the variation of CG properties as a function of the environment they inhabit and their assembly channel. 
Figure~\ref{fig:boxplot_h20} displays boxplot diagrams of the CG properties in H20 (similar figures for the other three SAMs are in Appendix~\ref{ap:boxplots}). The waists of the boxes represent the medians of the corresponding distributions, while notches in the boxes represent the confidence intervals for the medians. If the notches do not overlap, the medians are statistically different.
From top to bottom, we show the boxplots of the radial velocity dispersion\footnote{The radial velocity dispersion is calculated using the gapper estimator described by \cite{Beers+90}} ($\sigma_v$), the dimensionless crossing time\footnote{The dimensionless crossing time is computed as $H_0 t_{cr} =100 \pi h \langle d_{ij} \rangle / (2\sqrt{3}\sigma_v)$, where $\langle d_{ij} \rangle$ is the projected median inter-galaxy separation.} (${\rm H_0 t_{cr}}$), the r-band rest-frame absolute magnitude of the two brightest galaxy members ($M_1$ and $M_2$), and the difference between them ($M_2 - M_1$). 
The left column displays the CG properties for the total sample split according to the environment they inhabit. The far right column shows the same sample but split as a function of their assembly channels. The middle columns display the properties of CGs with a given assembly channel (from Early to Late) split according to the environments. 

First, we compare the results for CGs in SAMs (first column) with those obtained by \cite{taverna+23} for CGs in the SDSS DR16 (see their figure 5).
All the SAMs results for $\sigma_v$, ${\rm H_0 t_{cr}}$ and $M_2 - M_1$ reproduce well the tendencies found in observations. 
This is, CGs in the densest environments show larger $\sigma_v$ and smaller crossing time, while the opposite is observed in non-embedded systems, and the magnitude gap between the two brightest galaxies shows no variation across environments.
However, we notice that the median velocity dispersions obtained for CGs in A21 (280 - 400 km/s) are higher than the values obtained for the other SAMs (200 - 300 km/s). The observed median velocity dispersions of CGs in SDSS observed by \cite{taverna+23} are more in agreement with H20, G11 and G13.
Finally, \cite{taverna+23} reported that the first-ranked galaxies in CGs appear brighter when embedded in larger galaxy groups compared to non-embedded CGs. This result is recovered in three of the four SAMs analysed in this work (G11, G13 and A21), but H20 does not exhibit this trend.

Regarding the velocity dispersions, the orange solid dots within the blue boxes in the top left panels show the median velocity dispersion of the galaxy systems that host CGs. Regardless of the SAM, the median velocity dispersion in CGs embedded in large systems is between $50 - 120 \, \rm km/s$ (depending on the SAM) higher than the values obtained for the galaxy systems in which they are immersed. 
In addition, for most of the SAMs (H20, G11 and G13), the median velocity dispersion of the host galaxy systems is higher than the median velocity dispersions of CGs embedded in Filaments, Voids or non-embedded CGs. This might be an indication that the higher velocity dispersion in CGs embedded in galaxy systems could be a sign of correlation with their host environment. 
However, it is not the same for A21, where the median velocity dispersion of the host galaxy groups is consistent with the measurement obtained for CGs embedded in Filaments, Voids or non-embedded CGs. In this case, the higher values obtained for CGs embedded in galaxy systems cannot be attributed only to the environment where they are immersed. 
 
In \citetalias{DiazGimenez+21} they showed, regardless of the SAM, a dependence of crossing time and magnitude gap on assembly channel, while velocity dispersion decreased slightly from Early to Late CGs\footnote{The only SAMs in common with that work are G11 and G13. Also, the current samples are not the same since in the previous work only CGs with four members were considered.}. In this work, the last column in Fig.~\ref{fig:boxplot_h20} shows similar trends for crossing time and magnitude gap. Still, we also find a clearer dependence of velocity dispersions on the assembly channel (100 $km \ s^{-1}$ total variation for extreme median values) than in observations. 
As the first-ranked galaxy's magnitude in CGs was not considered in \citetalias{DiazGimenez+21}, we cannot compare our results with their work. In this work, we observe a tendency for brighter first-ranked galaxies in Early CGs than in Late CGs, only in G11 and A21 SAMs, and a constant trend in H20 and G13. 

Analysing how the properties of CGs behave based on both the environment and the assembly channel, from middle columns in Fig.~\ref{fig:boxplot_h20},  we observe that the general result observed for Late CGs in the right column, being typically characterised by low $\sigma_v$ values, do not longer hold when considering only Late CGs that are embedded in larger galaxy groups. These Late systems exhibit larger velocity dispersions quite similar to Early systems inhabiting groups. 
The fact that the total Late CG samples have relatively low $\sigma_v$ values, in general, is because these samples are dominated by non-embedded CGs (see Fig.~\ref{fig:barplot}). 
The results for Early CGs are not that straightforward. Early CGs are dominated by CGs embedded in larger galaxy groups (Fig.~\ref{fig:barplot}), which implies they have high values of $\sigma_v$ ($\sim$ 370 km/s, averaging over SAMs).
However, the $\sigma_v$ observed for the Early CGs that are non-embedded CG ($\sim$ 280 km/s) is typically higher than the observed for the whole sample of non-embedded CGs ($\sim$ 200 km/s). Hence, Early CGs display the highest velocity dispersion values regardless of the environment.
When comparing the velocity dispersions of the CGs embedded in galaxy systems split by assembly channels,  we observe that most of the CG embedded in galaxy systems exhibit larger median velocity dispersions than their hosts, regardless of the assembly channel, with the exceptions of Early CG embedded in groups in H20 and G13 where the velocity dispersions of the CGs are statistically indistinguishable to the $\sigma_v$ of their hosts. In these two SAMs, the more evolved systems tend to be more dynamically coupled to their hosts.

When considering the crossing times, the analysis follows the described above for velocity dispersions. 
Late CGs tend to have larger ${\rm H_0 t_{cr}}$ values (because non-embedded systems dominate them) than Early CGs. However, those Late CGs that are embedded in large galaxy groups can actually have shorter ${\rm H_0 t_{cr}}$ values than expected from the total sample of Late CGs.
On the other hand, Early CGs typically have the smallest ${\rm H_0 t_{cr}}$ values regardless of their environment, which suggests more frequent galaxy interactions during their longer lifetime than the Late CGs.

From the distribution of the absolute magnitude of the first-ranked galaxy in CGs, a very small dependence on the environment within each assembly channel is found for G11, G13 and A21. In these SAMs, the tendency of a brighter first-ranked galaxy in embedded systems observed in the whole sample is recovered also within each assembly channel. Still, there is no dependence either on the environment or the assembly channel for H20.

In addition, the middle panels of Fig.~\ref{fig:boxplot_h20} indicate that
the magnitude gap of CGs within a given assembly channel resembles the general trend of the left panel, i.e., no dependence on the environment. Instead, when moving between panels (different assembly channels),
the first-ranked galaxy becomes more dominant the earlier the system is formed, regardless of the semi-analytic model used. Therefore, the magnitude gap in CGs is dependent on the assembly channel rather than the environment.This is a significant finding because the magnitude gap is an observable CG property that could be directly associated with their assembly history, independent of the environment. As a result, the magnitude gap is a suitable measurement for performing potential observational segregation on the assembly channel. 

Analysing the brightest galaxy in groups (BGG), \cite{shen+14} concluded that galaxy groups with a prominent and luminous BGG should be correlated with a large magnitude gap between the two brightest galaxies in their system. This may prompt the question of how it is possible that CGs exhibit a correlation between the magnitude gap and the assembly channel, yet there may not necessarily be a correlation between the brightness of the first-ranked galaxy and the formation history of the compact group. Our results show that there are several possible paths to build a dependence on the magnitude gap. Firstly, it is worth noticing that the strength of the dependence of the magnitude gap-assembly channel relation depends on the SAM (although it is present in all of them).
From the rightmost panels, it can be seen that H20 shows a less pronounced downward slope for the magnitude gap ($\sim$0.3 difference in the magnitude gap between Early and Late CGs), followed by A21 ($\sim$0.5), while G11 and G13 display the more pronounced slopes ($\sim$0.6).
From the magnitudes of the first and second-ranked galaxies in CGs (third row of Figures~\ref{fig:boxplot_h20},~\ref{fig:boxplots_g11},~\ref{fig:boxplots_g13} and \ref{fig:boxplots_a21}),
we observe for H20 an almost constant trend for $M_1$ and $M_2$ but with a tendency to have slightly brighter $M_1$ and slightly fainter $M_2$ for Early CGs than Late CGs, enough to build the magnitude gap dependence. 
For G13, a quite constant trend for $M_1$ and an increasing $M_2$ trend is observed, which implies that the magnitude gap is produced because the $M_2$ galaxies are fainter in Early CGs than their counterpart in Late CGs. 
For G11, a clear decreasing trend for $M_1$ and an increasing one for $M_2$ produce the trend for the magnitude gap when moving from Early to Late CGs. 
Finally, A21 displays a clear decreasing trend for $M_1$ and an almost constant trend for $M_2$, obtaining a magnitude gap trend that is a reflection of the first-ranked galaxy behaviour. Therefore, we conclude that the possible paths to build a decreasing magnitude gap as a function of the assembly channel in CGs are not limited exclusively to the brightening of the first-ranked galaxy with the age of the system.

\section{Summary and Conclusions}
\label{sec:conclusions}
In this work we follow the steps of \cite{taverna+23} to understand the environment surrounding compact groups (CGs) of galaxies in the Universe. We complement their observational study performed in CGs and expand it by performing an identical study but this time on CGs identified in mock catalogues constructed from four semi-analytical models (SAM) of galaxy formation applied on top of a Millennium N-body cosmological numerical simulation \citep{Springel+05, Guo+13}. The SAMs used in this work were built by \cite{Guo+11} (G11), \cite{Guo+13} (G13), \cite{Henriques+20} (H20) and \cite{Ayromlou21} (A21).

This work is the fifth in a series of studies about CGs identified in catalogues built with different SAMs. 
We aim to answer two trigger questions: 1) do CGs identified in different SAMs reproduce the environment surrounding CGs in observational catalogues?; 
2) following previous studies (\citetalias{DiazGimenez+21} and \citetalias{zandivarez+23}), we wonder if CGs characterised by an assembly channel are found preferentially in a particular large-scale environment.

Firstly, we observed that CGs identified in most of the SAMs closely reproduce the occupation rate of CGs as a function of the environment obtained in observational data by \cite{taverna+23}.
We confirm that roughly 50\% of CGs are non-embedded systems while $\sim 40\%$ are embedded in loose groups or nodes of filaments. The remaining CGs ($\sim 10\%$) can be found almost equally distributed in filaments and cosmic voids. 
Overall, these results are consistent with the observations throughout three different cosmological models and several different sets of recipes to model the physical processes that govern galaxy formation.   

Secondly, we implemented the classification of CGs by their assembly channel introduced in \citetalias{DiazGimenez+21}. In this case, we extend the criteria to CGs with three and four galaxy members which represents more than 90\% of the whole sample of CGs. 
Given the four possible channels of assembly, which range from recently to early formed (Late, Second, Gradual and Early channels), we find that the majority of CGs can be classified as Late assembled ($\sim 43\%$) and a large fraction as being in their Second pericenter passage ($\sim 24\%$). Therefore, two-thirds of CGs can be considered as recently assembled. Roughly 15\% of CGs are consistent with a gradual contraction formation scenario (Gradual), while very few ($\sim 5\%$) have a very early assembly (Early). The rest of the CGs are false detections or chance alignments in the line of sight during the identification process. 

When these two characteristics, environment and assembly channel, are combined, interesting results are observed. Although there is not a clear one-to-one correspondence between an assembly channel and a specific environment, there is a tendency for CGs that have assembled in early times to be predominantly embedded in larger galaxy systems ($\sim 60\%$), while approximately a third of the early assembled groups is found non-embedded. This result contrasts with what occurs for CGs that have recently assembled, that are predominantly non-embedded systems, and only a third inhabit larger galaxy groups.  

The combined effects of environment and assembly channels were studied on the properties of CGs, raising questions about the old nature vs. nurture paradigm. 
In general, recently formed CGs are characterised by lower velocity dispersions and higher crossing times. 
This is primarily because these systems are predominantly non-embedded, and it has already been demonstrated that is directly related to the environment \citep{taverna+23}. However, there is a subset of recently formed CGs that are embedded in larger systems that do not exhibit the same trends in these physical properties as the total sample. These late-forming embedded systems have relatively high-velocity dispersion and low crossing times, and this is more characteristic of the environment in which they inhabit than of their formation channel. This dependence on the environment is not as prominent for the early-forming CGs; in general, they exhibit the highest velocity dispersions and the lowest crossing times regardless of the environment. 
These physical features of the early-formed CGs have a great impact on the evolution of their galaxy members. The very short crossing times signature for these systems may indicate that galaxy interactions are prone to be quite frequent within them during their long lifetime, having an important impact on the galaxy properties evolution. In \citetalias{zandivarez+23} we have demonstrated that early-formed CGs have more extreme losses of gas content, faster growth rate in the mass of black holes and more marked galaxy quenching as a function of cosmic time than the observed for recently-formed CGs. Even though various mechanisms may be responsible for these qualities, its long evolution in an extremely confined volume of space favouring tidal interactions is one of the fundamental ingredients to achieve such extreme galactic evolution.

Lastly, among the analysed properties, one of the most intriguing is the magnitude gap between their two brightest galaxies. In \citetalias{DiazGimenez+21}, it was noticed that the magnitude gap might show a variation with the assembly channel, while \cite{taverna+23} reported that CGs in the SDSS do not present a variation of the magnitude gap with the environment. In this work, we confirmed that the magnitude difference between the two brightest galaxies in CGs does not vary with the environment but rather only changes with the assembly channel. Hence, the dominance of the brightest galaxy in the CG could be a potential proxy for establishing selection criteria for early-forming compact groups in observational catalogues. 
This relation between a larger magnitude gap in early-formed CGs resembles the classical definition of fossil groups \citep{jones03}. These systems have been studied largely with both observational (e.g., \citealt{Khosroshahi+07,santos+07,zarattini+15}) and numerical data (e.g., \citealt{DiazGimenez+08,Dariush+10,Kanagusuku+16}) showing that the large magnitude gap is an indication of older galaxy systems, with high concentrated halos and central first-ranked galaxies brighter than their counterpart in galaxy systems with large magnitude gap. Our results show that early-formed CGs are characterized as dense galaxy systems with a dominant first-ranked galaxy, although this dominance is not necessarily associated with a first-ranked galaxy brighter than their counterpart in recently-formed CGs. 

About the influence of the different SAM recipes used in this work, we noticed a good agreement between the G11, G13 and H20 SAMs. 
Even when A21 SAM closely reproduces the frequencies of CGs as well as the overall trends reported as a function of the environment and the assembly channels, there are some differences that are worth mentioning. 
About the percentage of CGs with the environments, A21 SAM shows some differences since produces $\sim 7\%$ fewer CGs in Nodes and an excess of $\sim 10\%$ of Non-embedded CGs. Also, A21 SAM displays $\sim 8\%$ fewer Early+Gradual CGs and a $\sim 5\%$ excess of Fake CG systems. 
When comparing with the median velocity dispersions obtained for CGs in G11, G13 and H20, A21 shows a larger median velocity dispersion regardless of the environment. In \citetalias{Taverna+22}, it was observed that A21 showed the lowest galaxy and CG density compared with other SAMs. In \citetalias{zandivarez+23}, the A21 differs from the other SAMs for having the lowest fraction of orphan galaxies per bin of stellar mass as well as the largest fraction of satellites. This previous work also reported that when analysing the suppression of star formation in galaxies within CGs under a similar cosmology (A21 and H20), the model of A21 showed the largest fractions of quenched galaxies. Clearly, the implementation of semianalytical recipes in A21 is critical to determining the resulting abundance of CGs and the properties of their galaxy members. The stronger supernova reheating and ejection efficiency as well as the extension of tidal stripping to all satellite galaxies in A21 are probably some of the main reasons to account for the differences observed with the other SAMs.

Finally, our findings might place us in a scenario where CGs embedded in larger, overdense systems can survive longer, evolving while maintaining their compact state for more than half the age of the Universe. One can think as if the galaxy system containing them somewhat acts as a shield against factors that could potentially disturb the evolution of the compact group and dismantle it. 
Clearly, these CGs are more evolved systems: they exhibit higher velocity dispersions and relatively small sizes, as well as more evolved central galaxies in terms of their dominance in luminosity compared to their closer brighter companions. 
While it is true that early-forming CGs not embedded in larger structures also exist, they are less likely. Nonetheless, early-formed non-embedded CGs also show signs of being more evolved (with similar characteristics to the embedded ones), despite not inhabiting a larger and relaxed overdensity. Therefore, although less likely, early assembled CGs are able to survive and evolve in the most varied environments in the universe.  

An interesting topic to be addressed could be how CGs are related to the different types of assembly bias. In \citetalias{zandivarez+23} we have demonstrated that CGs can be considered a new laboratory for studying the \emph{galaxy assembly bias} since, at a given stellar mass, galaxy members of early-formed CGs actually formed their stars also earlier. 
Our current research shows that CGs formed early tend to be found within larger galaxy systems. This raises 
the question of whether the systems hosting the CGs are also early-formed, and if so, whether these CGs can be used to explore the \emph{halo assembly bias}. 
However, this possibility is not straightforward since it might also depend on when the Early CGs were embedded into their hosts.  
We plan to delve further into this intriguing topic in future research.

\section*{Acknowledgements}
We thank the anonymous referee for useful comments that improved the final version of our work.
We thank the authors of the SAMs for making their models publicly available. 
The Millennium Simulation databases used in this paper and the web application providing online access to them were constructed as part of the activities of the German Astrophysical Virtual Observatory (GAVO).  
This work has been partially supported by Consejo Nacional de Investigaciones Cient\'\i ficas y T\'ecnicas de la Rep\'ublica Argentina (CONICET) and the Secretar\'\i a de Ciencia y Tecnolog\'\i a de la Universidad de C\'ordoba (SeCyT).

\section*{Data Availability}
The data underlying this article were accessed from \url{http://gavo.mpa-garching.mpg.de/Millennium/}. Galaxy lightcones built-in Paper~I and used in this work are accessed from \url{https://doi.org/10.7910/DVN/WGOPCO} \citep{data_paperI}.
The derived data generated in this research will be shared with the corresponding authors at reasonable request.



\bibliographystyle{mnras}
\bibliography{refs} 




\appendix

\section{CG Physical Properties}
\label{ap:boxplots}

\begin{figure*}
    \begin{center}
    \includegraphics[width=0.90\hsize]{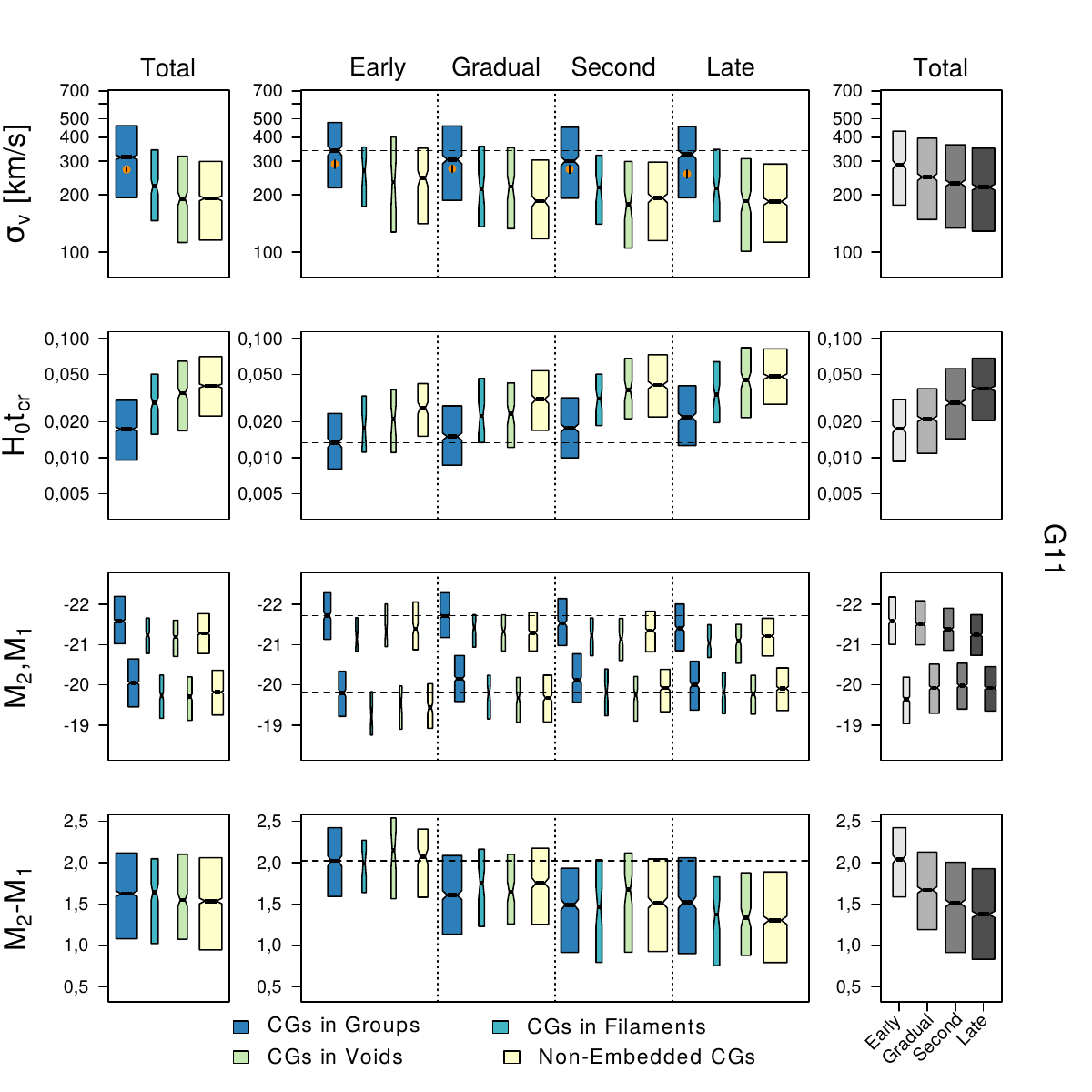}
    \caption{\label{fig:boxplots_g11} Same as Figure \ref{fig:boxplot_h20}, but for the CGs identified in the semi-analytical model G11.
    }
    \end{center}
\end{figure*}

\begin{figure*}
    \begin{center}
    \includegraphics[width=0.90\hsize]{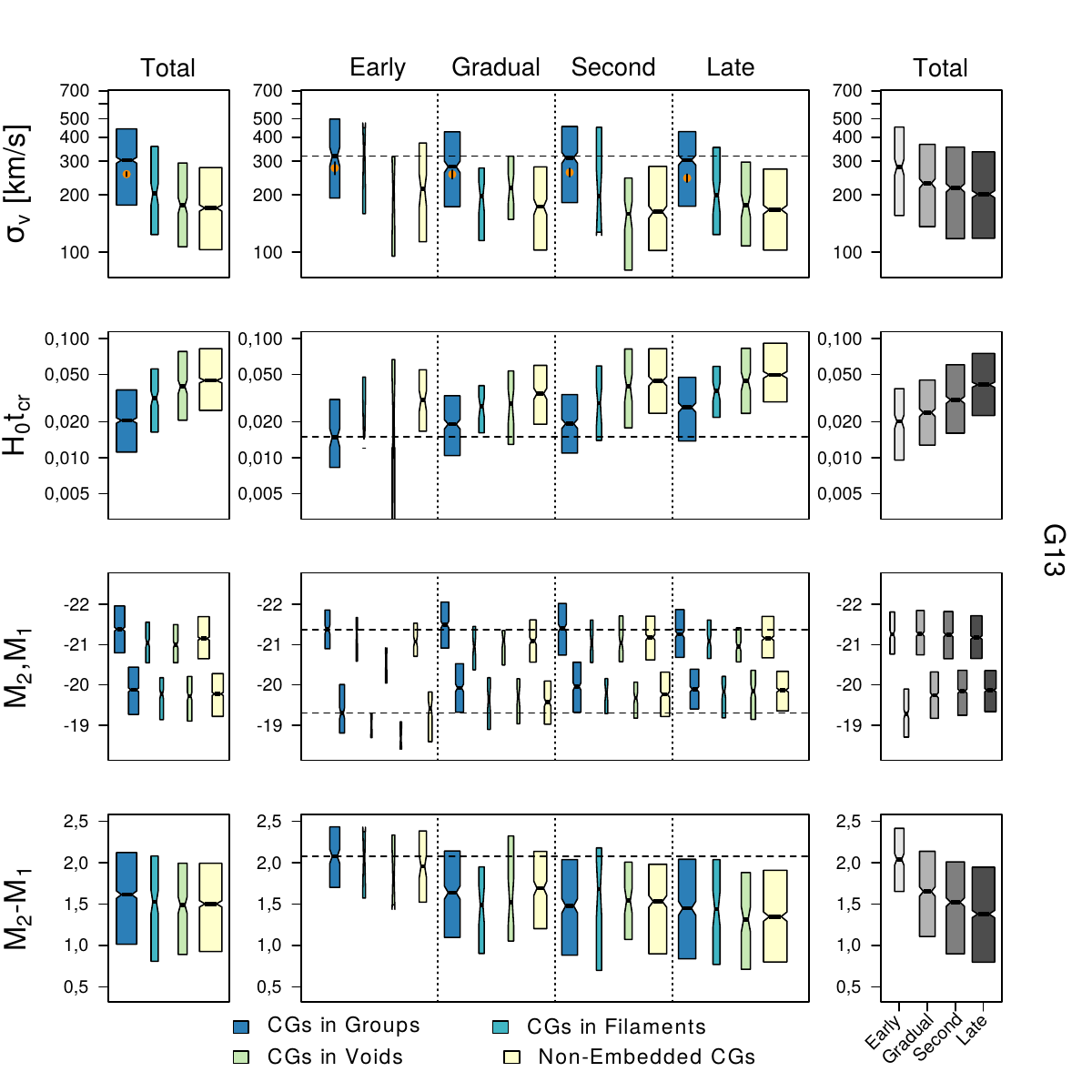}
     \caption{\label{fig:boxplots_g13} Same as Figure \ref{fig:boxplot_h20}, but for the CGs identified in the semi-analytical model G13.
    }
    \end{center}
\end{figure*}

\begin{figure*}
    \begin{center}
    \includegraphics[width=0.90\hsize]{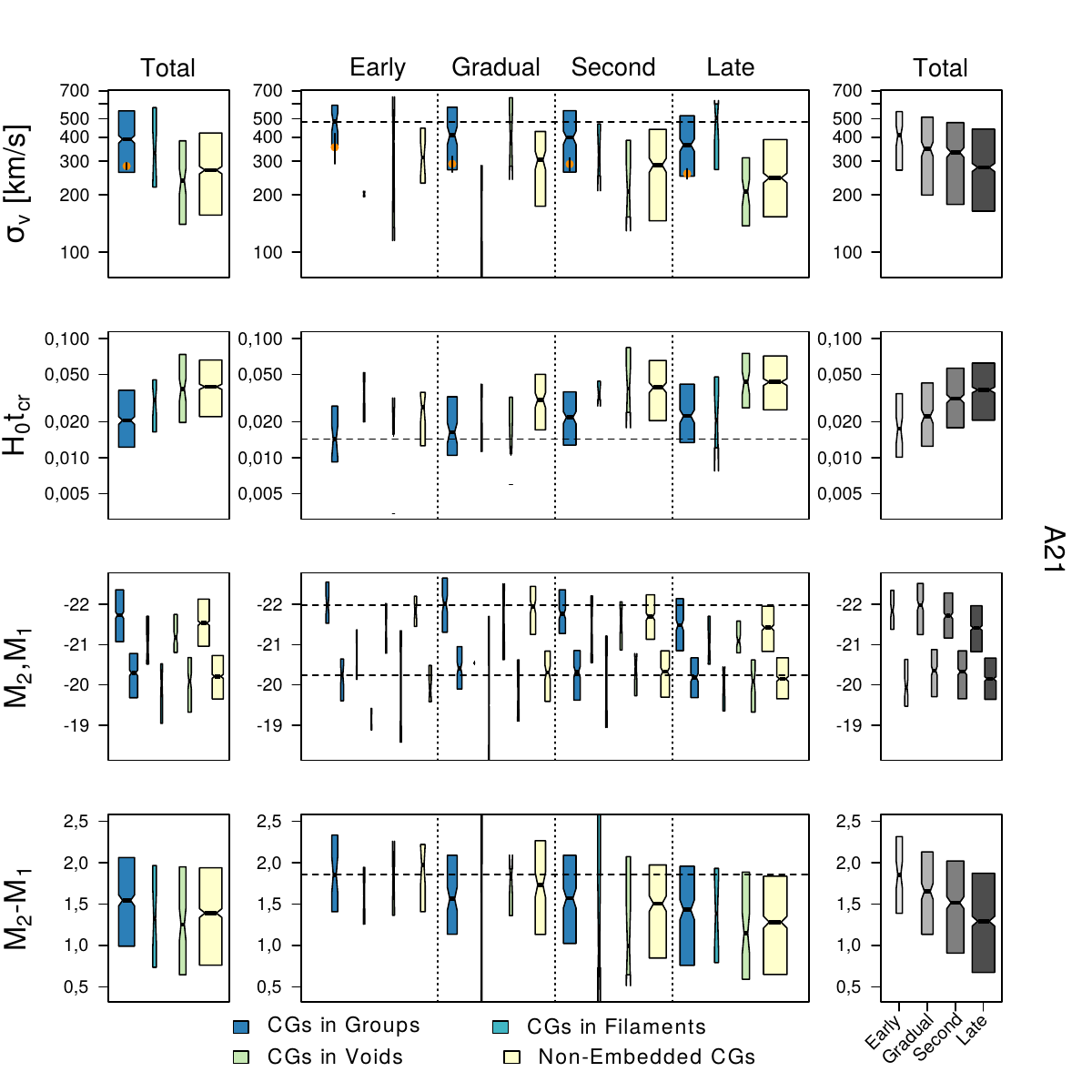}
     \caption{\label{fig:boxplots_a21} Same as Figure \ref{fig:boxplot_h20}, but for the CGs identified in the semi-analytical model A21.
    }
    \end{center}
\end{figure*}

In Figures~\ref{fig:boxplots_g11}, \ref{fig:boxplots_g13}, and \ref{fig:boxplots_a21}, we show the boxplots diagrams of CG physical properties for the SAMs G11, G13 and A21, respectively. 

\bsp	
\label{lastpage}
\end{document}